\documentclass[a4paper,12pt]{article}
\usepackage[T1]{fontenc}                  
\usepackage[utf8]{inputenc}               
\usepackage{textcomp}
\usepackage{lmodern}                      
\usepackage{setspace}                     
\usepackage{sectsty}                      
\usepackage{natbib}                       
\usepackage[protrusion=true,expansion=true]{microtype} 
\usepackage{booktabs}                     
\usepackage{graphicx}                     
\usepackage{amsmath}                      
\usepackage{rotating}                     
\usepackage{latexsym}                     
\usepackage{bm,array}                     
\usepackage[tableposition=top,labelfont=bf]{caption}   
\usepackage{authblk,etoolbox}             
\usepackage{epstopdf}                     
\usepackage{parskip}                      

\makeatletter
\patchcmd{\@maketitle}{center}{flushleft}{}{}
\patchcmd{\@maketitle}{center}{flushleft}{}{}
\def\maketitle{{%
  
  \AB@maketitle}}
\makeatother

\DeclareUnicodeCharacter{00A0}{ }

\title{\textbf{Toward a Unified Understanding of Casualty Distributions in Human Conflict}}
\author{Michael Spagat, Royal Holloway University of London
\and Stijn van Weezel, Radboud University
\and Minzhang Zheng, Michigan State University
\and Neil F. Johnson, George Washington University}
\date{ }

\begin{document}
\maketitle
\noindent

\begin{abstract}
\small{\bf{We are able to unify various disparate claims and results in the literature, that stand in the way of a unified description and understanding of human conflict. First, we provide a reconciliation of  the numerically different exponent values for fatality distributions across entire wars and within single wars. Second, we explain how ignoring the details of how conflict datasets are compiled, can generate falsely negative evaluations from power-law distribution fitting. Third, we explain how a generative theory of human conflict is able to provide a quantitative explanation of how most observed casualty distributions follow approximate power-laws {\em and} how and why they deviate from it. In particular, it provides a unified mechanistic interpretation of the origin of these power-law deviations in terms of dynamical processes within the conflict. Combined, our findings strengthen the notion that a unified framework can be used to understand and quantitatively describe human conflict. 
}}  \end{abstract}

One of the most remarkable results in conflict research is that the distribution of the aggregate numbers of fatalities in entire wars (i.e. the distribution of war sizes) can be well modeled as an approximate power-law \citep{cederman2003modeling,gonzalez2016war}.
This empirical regularity is known as Richardson's Law --- after polymath Lewis Fry Richardson who first studied this phenomenon more than half a century ago \citep{richardson1948variation,richardson1960statistics}.
Approximate power laws also tend to capture well the distributions of event sizes, measured by fatality counts, within wars \citep{bohorquez2009common,johnson2013simple,spagat2018fundamental} and terrorism \citep{clauset2007frequency,spagat2018fundamental} and generative models have been suggested \citep{clauset2007frequency, bohorquez2009common}. 
Here we address some unresolved issues and misunderstandings that are inhibiting a more unified view of this subject. 

Our paper makes three contributions to a more unified understanding of  human conflict. First, we show that the approximate power laws obtained across wars, treating the number of fatalities in each war as a single data point, bear a simple relationship to the approximate power laws found within single wars.  
Specifically we show through simulations that the ranges of power-law exponents found within individual conflicts will, when we aggregate the event data they generate into complete wars, produce the sorts of power-law exponents researchers have found across whole wars (Fig. 1). 
To our knowledge, this is a novel insight that opens new opportunities for a unified understanding of war from both microscopic and macroscopic points of view.  Thus, the research program initiated by Richardson many decades ago seems to be leading to a synthesis of micro and macro elements that is consistent with the spirit of his life's work, including his research on weather and fractals \citep{gleditsch2020lewis}.
Second, we highlight a common feature of conflict event data that is often ignored, leading some researchers to exaggerate evidence against power laws in some conflicts (Fig. 2). 
Much conflict event data contains a mixture of pure events with what we term "composite events" which are single data points that aggregate multiple pure events into a single "event" with a single fatality number. 
Ignoring this data reality leads to biased estimation. We do not claim exact power laws everywhere in within-war event data:  
sometimes power laws are relatively poor fits for conflict event data \citep{spagat2018fundamental}. But with respect to this, our third point is that our previously introduced generative theory of human conflict \citep{bohorquez2009common} does not assume pure power-law behavior. Instead, (i) our generative theory of warring populations \citep{bohorquez2009common} describes and explains approximate power-laws when they exist, but also (ii) it explains conflict data where there are departures from power laws, and (iii) it explains these departures in terms of microscopic dynamical processes (Fig. 3).

\section{Connecting Single Wars to All Wars}
We have found an important connection between estimated power laws that fit the distribution of whole war sizes, and the range of estimated power laws that fit event sizes within individual wars. 
The estimated exponents for the latter are generally higher than those for the former -- however we have shown through simulations that the process of aggregating within-war fatality counts into totals for whole wars is able to reconcile these differences. 
Thus, it now appears possible to unify the two branches of the literature on war and power laws. 

\citet{cederman2003modeling}, \citet{clauset2018trends} and most recently \citet{braumoeller2019only}, build on Richardson's seminal work, showing that the distribution of severities for entire wars is an approximate power-law within an estimated range of  $\alpha\sim 1.5-1.7$, depending on the date range and war types included. 
The goodness-of-fit values seem to be fairly high as well ($p\sim 0.5$). 
As noted above, we find that events within each \textit{individual} war $i$, with event sizes measured by fatalities, show an approximate power-law distribution which is spread broadly around an exponent value of $\beta_i\sim 2.5$, again with reasonably high $p$ values. 
The power-law testing procedure is the same in both cases: in the latter case of an individual war $i$, the process involves taking the histogram of the severity of individual events within this war and performing the standard power-law test to get the best-fit power law exponent (which we will refer to as $\beta_i$). 
In the former case of entire wars, the process involves taking the histogram of the severity for each entire war $W_1, W_2,...$ etc. and then getting the best-fit power law exponent (which we will refer to as $\alpha$). 

\begin{figure}[!ht]\centering
  \includegraphics[scale=.9]{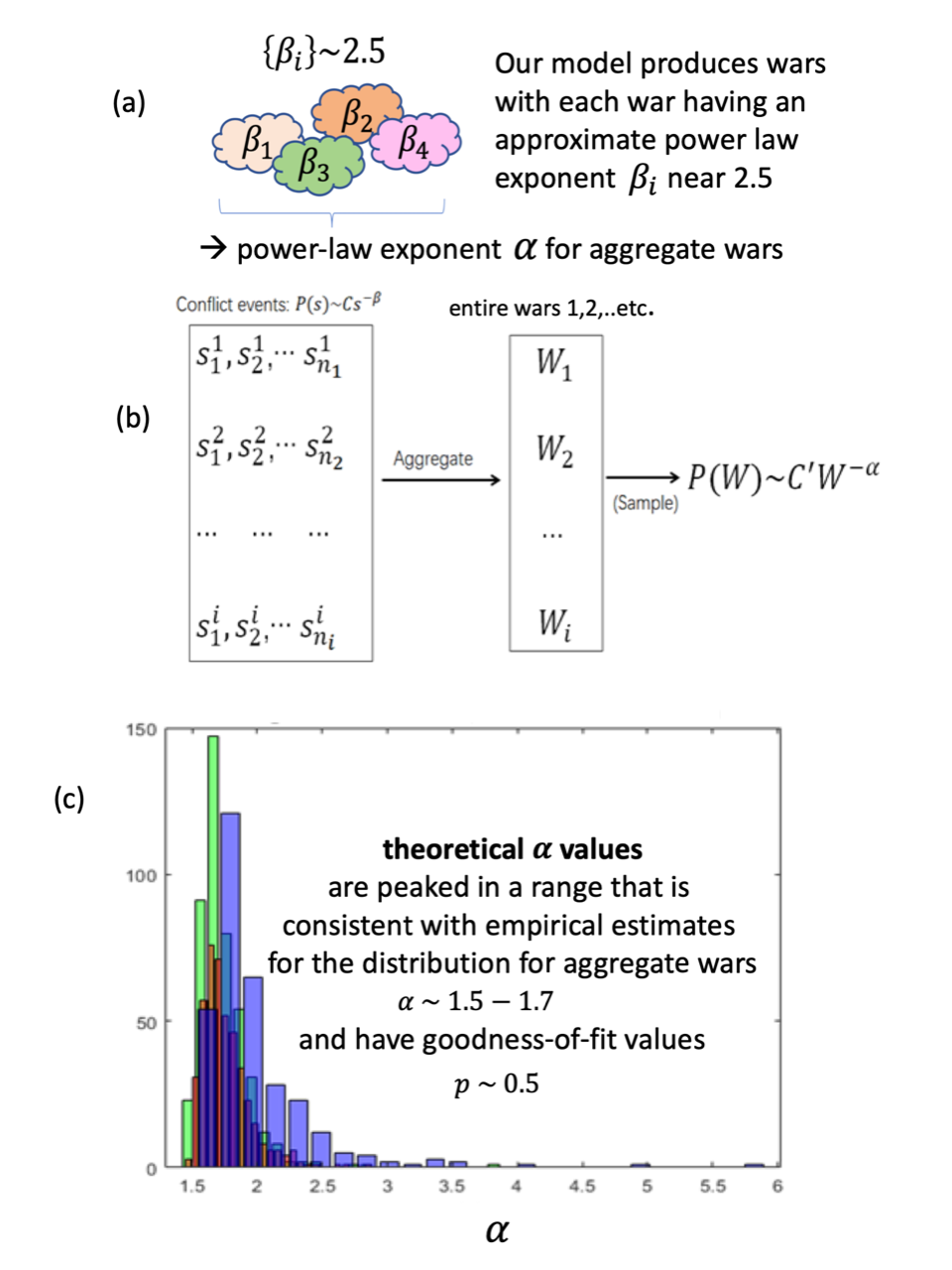}
  \caption{(a) Schematic of the simulation procedure, shown in more detail in (b). To generate a number of wars consistent with our model, we generate events from power-law distributions with exponents distributed around 2.5 (see text). The aggregate size of each war is calculated, yielding $\{W_i\}$. This set of $\{W_i\}$ values is then sampled to mimic the process of using a database of known wars with known aggregate totals.  As shown in (c), the resulting power law exponents for the aggregate size of entire wars, is in the range observed empirically. Each distribution is for a different number of mean events per war. In each case, the distribution of $\alpha$ values tends to be peaked in a range that is consistent with the empirical values for entire wars (i.e. $\alpha\sim 1.5-1.7$) and they each have reasonably high goodness-of-fit values $p$ ($p\sim 0.5$).}
  \label{fig:results}
\end{figure}

The question therefore arises: are these two findings for the power-law test for (i) aggregate wars yielding exponent  $\alpha\sim 1.5-1.7$ and (ii) individual wars yielding exponents $\{\beta_i\}\sim 2.5$, consistent with each other? 
Below we show through simulations that they are indeed consistent. 
Thus, we provide the first unified treatment of Richardson's law, crossing the boundary separating fatality events within individual wars from aggregated fatality totals across entire wars. 
This connection opens a new chapter in the research program initiated by Richardson many decades ago. 

To show this, we first simulate a set of individual events for an individual war $i$ guided by our model. 
Though we could in principle generate these events using the cluster interactions from our model discussed in Sec. 3 and Fig.~\ref{fig:extra}, or the more basic one-dimensional version, for simplicity for a given war $i$ we instead choose to generate random events from a power-law with exponent $\beta_i$, where $\beta_i$ is picked randomly from a distribution spread around $2.5$. 
We then repeat this procedure to generate a number of different simulated wars, allowing a different number of events $n$ for each war. 
We have checked the robustness of our results for different choices of peaked distribution for these $\beta_i$ values, and also different choices of $x_{\rm min}$ for the power-law onset. 
Here for simplicity, we show results for a normal distribution of $\beta_i$ values. 
We have also checked that our results are robust to different choices of standard deviation in $\beta_i$ around $2.5$. 
We have also checked the robustness for having a peak away from $2.5$, and we have checked the robustness to having different numbers of events for each simulated model war $i$.  

In all cases, we find that the results that we provide in Fig.~\ref{fig:results} are indeed representative and the core finding is robust. 
The aggregated casualty total for each war $i$ is given by $W_i$ which is the sum of the individual events $1,2,..$ etc. within that war, i.e. $W_i=s_1^i+s_2^i+...$. 
This is shown schematically in Figs.~\ref{fig:results}(a) and (b). 
This entire process provides us with a set of values $\{W_i\}$ corresponding to the total fatalities in our simulated model wars $i=1,2,..$. 
This represents our model's predicted record for all human wars. 
In the real world, only some prior wars have data available for them. 
Indeed, some wars may have been lost from the history books and hence their existence is unknown. 
We therefore sample subsets of $\{W_i\}$ in order to mimic the known history books, and hence mimic the existing database of wars analysed by other researchers. We then proceed as if with real data, by doing the usual power-law test. 

Fig.~\ref{fig:results}(c) shows the striking result that the resulting distribution of entire war exponents for different samples tends to be bunched in the same range $\alpha\sim 1.5-1.7$ as in the empirical findings. 
Moreover, the goodness-of-fit values are distributed around $p\sim 0.5$. 
We chose for simplicity $x_{\rm min}=1$, and the $\{\beta_i\}$ to be distributed normally with a mean at $2.5$ and a standard deviation of $0.5$, similar to the empirical findings in our previous work. 
We also chose the number of events per war to follow a lognormal distribution which is again consistent with actual war data, and we show the resulting ${\alpha}$ distributions for three representative values of the mean in Fig.~\ref{fig:results}(c). 
We also chose $1000$ total wars $\{W_i\}$, and sampled subsets of size $100$. 
Again, we checked that the results in Fig.~\ref{fig:results} are robust to variations in each of these choices. 

This finding unifies the power-law estimation and testing results for event sizes within individual wars with the corresponding results for fatalities aggregated over entire wars, capturing how the scaling coefficient changes as we move from the intra-conflict level to the inter-conflict level. 
This means that looking at individual violent events within a single war is not the same as looking at individual wars within a collection of many wars, despite the fact that both phenomena can be captured reasonably well by power laws -- albeit different ones. 
This observation may seem obvious to a historian but it is interesting to see it emerge for the first time from realistically simulated data while giving us additional insights into how this difference can be quantified. 
In particular, we see how compiling aggregate data across wars has the impact of lowering the value of the best-fit exponent. 
But why does it happen? 
Our simulations show that this is because in the production of the simulated wars, there is always a reasonable chance to generate a war with both a relatively high ratio of large events to small events \textit{and} and also a large total number of events overall.  
These are the really big wars (i.e. large $W$) that flatten out the power law curve. 
This introduction of high $W$ values tends to increase the fatness of the tail for the aggregated wars, and hence lower the value of the overall power-law exponent below $2.5$ -- as observed in Fig.~\ref{fig:results}(c).

\section{Estimates of Power Law Goodness-of-Fit}
Estimates of goodness-of-fit for power-laws based on publicly available datasets viewed as pure 'event data' can be misleading and wrong. 
This is because many such datasets mix together true events with composite events.
For example, the IBC (Iraq Body Count) database has entries taking the following form

\begin{quote}
  \textit{X bodies delivered to the Baghdad morgue in the month of M in the year of Y.}
\end{quote}

Such entries are not single discrete events in which $X$ \textit{people} were killed. 
In fact, such an entry may be closer to a composite of $X$ \textit{events} in which 1 person is killed. 
Moreover, IBC is not unique in including both true and composite events. 
Other databases also contain entries that are composites of multiple events. 
For example, the commonly used GED (Georeferenced Event Database) curated by the Uppsala Conflict Data Program (UCDP), has many such entries.  
Fortunately, these are flagged in the dataset with a special indicator variable, (\texttt{event\_clarity}), so they are easy to exclude.
About 10\% of the observations in this dataset are composite events. 
This is not a huge number but the treatment of these events has a noticeable effect on inference.   

The civil war in Angola provides a useful illustration of the impact of composite events which constitute 36 percent of the Angola events in the latest version of the GED (v.19.1).  
Including all events --- regardless of whether they are true events or composites --- leads to an estimated $p$-value (i.e. goodness-of-fit) of 0.14 using the power-law test in R's `poweRlaw' package \citep{gillespie2015fitting}.
This value balloons up to 0.81 when we exclude composites.  
Mixing the composite events into the estimation and testing procedures doesn't pull the $p$-value for Angola down to a standard rejection threshold -- however the impact on the evidence against the power law hypothesis is both large and spurious.

\begin{figure}[!ht]\centering
  \includegraphics[scale=.7]{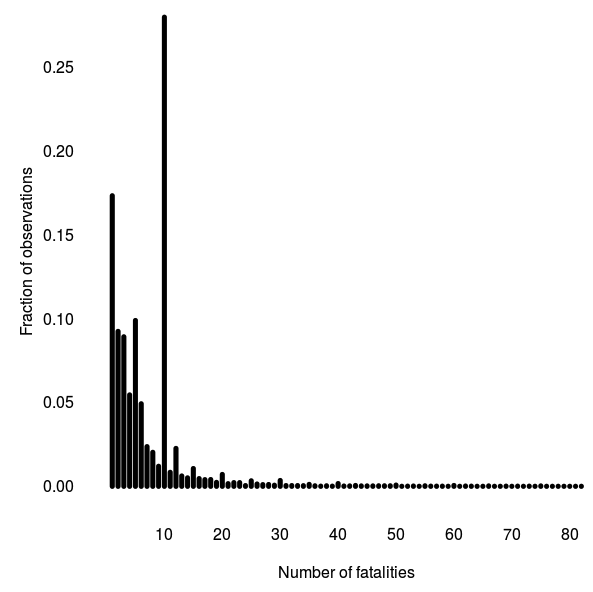}
  \caption{The fraction of observations with number of fatalities $x$ for all $x > 0$ for the conflict in Yemen (2015--19). 
  \newline \textit{Data source:} ACLED}
  \label{fig:yemen}
\end{figure}

\citet{zwetsloot2018testing} also tests for power laws using the ACLED (Armed Conflict Location \& Event Data Project) which is even less appropriate for this purpose than is the GED database if one does not first purge the composite events. 
ACLED has the practice of simply coding 10 deaths whenever the real number of people killed in an event is unknown, thereby creating a database that mixes together true events with many other that are, essentially, fabricated 10's.  
Size distributions of events based on the ACLED data show huge and artificial spikes at 10 for many countries such as Yemen (Fig.~\ref{fig:yemen}).
Again, naive application of power-law software to raw ACLED data will produce inaccurate results.   

\section{Generative Theory Beyond Power Laws}
\begin{figure}[!ht]\centering
  \includegraphics[scale=.55]{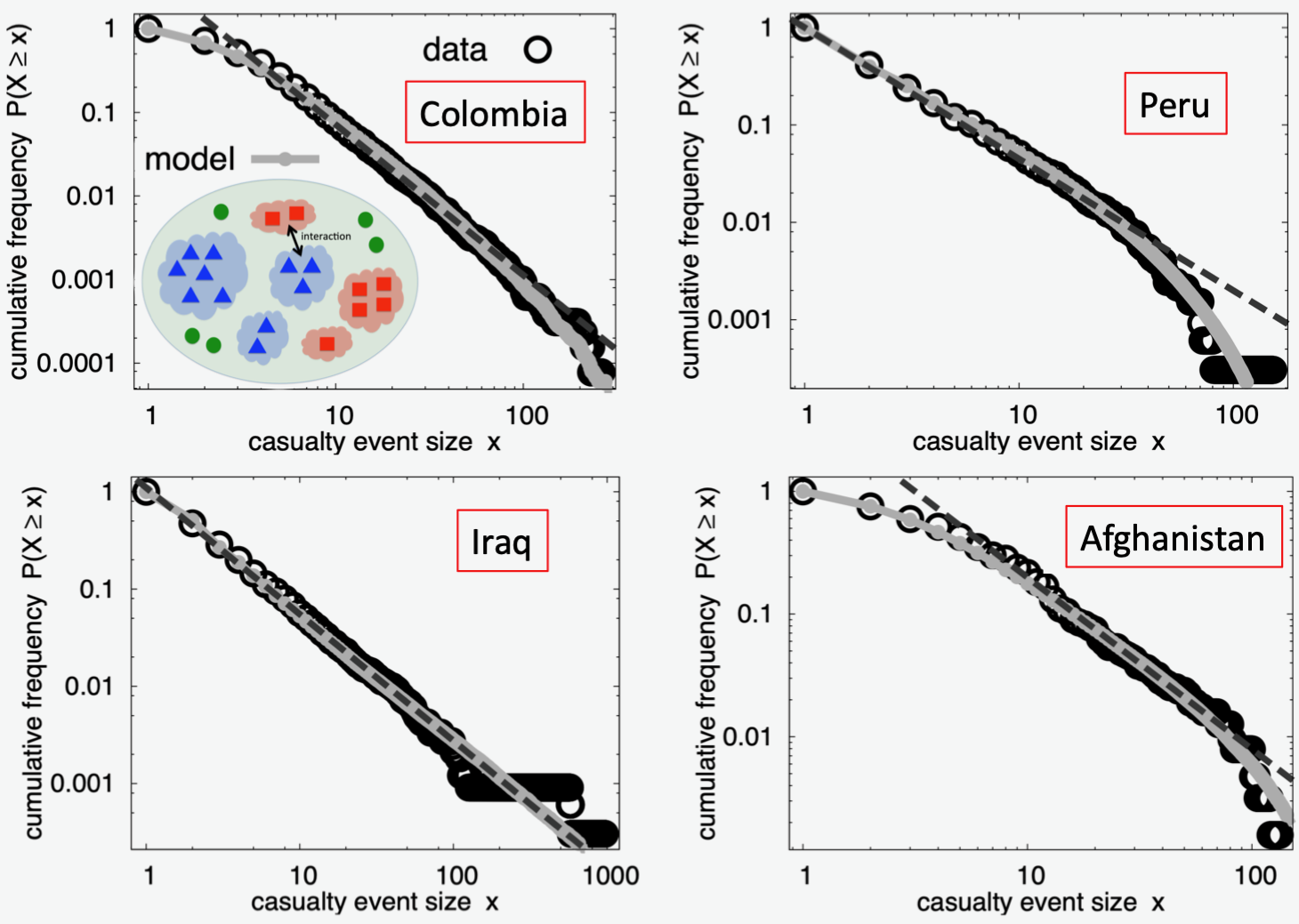}
  \caption{This plot shows the complementary cumulative distribution of event size $P(X \geq x)$ (i.e. the fraction of events having greater than or equal to $x$ casualties) for four conflicts, on a log-log scale. 
  It is adapted from \citet{bohorquez2009common}. 
  The horizontal axis shows the number of casualties in an event, i.e. event size $x$.
  The solid curves show the results from our model. 
  Our model is shown in the inset, and represents a two-population Red-Blue conflict (clusters of blue triangles and clusters of red squares that interact) with Greens (green circles) being the civilians. 
  The dashed line is simply a straight line guide-to-the-eye, not a power-law fit.}
  \label{fig:extra}
\end{figure}

Nobody that we are aware of, claims that conflict event sizes literally follow power laws, although \citet{zwetsloot2018testing} appears to characterize us as making this claim. 
It does turn out, however, that large swathes of conflict event data are well fit by power-laws with exponents with values clustering around 2.5 \citep{bohorquez2009common,johnson2013simple,spagat2018fundamental}, having $p$-values for a power law hypothesis that are well above standard rejection thresholds. 
Yet we have always noted dating back to \citet{bohorquez2009common} that there are important deviations from this approximate power-law pattern and we have worked to reveal and explain these in our modelling. Indeed, 
\citet{bohorquez2009common} explicitly presents our main two-population model featuring conflict between Red and Blue populations (inset Fig.~\ref{fig:extra}).
In addition to capturing the approximate power-law shape and slope in the event size distributions, it also reproduces the conflict-dependent deviations {\em beyond} a power law. 
This is shown explicitly by the solid curves in Fig.~\ref{fig:extra}. 
Any claim that our model stands or falls on there being a power-law and/or an exponent $2.5$, is incorrect (e.g. \citet{zwetsloot2018testing}).

In \citet{bohorquez2009common} and \citet{johnson2013simple}, we then go on to discuss how a more basic one-population version of our model can be solved mathematically using calculus and without the need for any numerical simulation. 
Seeking such a mathematically tractable vanilla version of our model, though much more basic in terms of its features, makes sense as a way of trying to understand the role of the approximate 2.5 empirical exponent value and also for checking limiting cases of the simulations for our actual two-population model. 
We reported how this very basic one-population version predicts an approximate power-law with an exponent of 2.5, and hence provides a simple benchmark reference for discussing regularities and deviations across the spectrum of conflicts. 
We discuss this also in \citet{spagat2018fundamental}. 
Specifically, the simpler one-population version is obtained by replacing the impact of the Blue population by a probability of Red cluster fragmentation. 
However, we stress that our full model --- i.e. the two-population version of the model --- goes much further by also explaining the deviations at low and higher casualty numbers, and hence the deviations from a power-law as shown in Fig.~\ref{fig:extra}. 
Broadly speaking, our model predicts that the ratio between the two populations' strengths (Red and Blue) tends to control the general behaviour of the slope, with greater strength differences resulting in steeper slopes, while the total Red strength tends to control the large high-end roll-off in Fig.~\ref{fig:extra} \citep{bohorquez2009common}. 
The bottom line is therefore that the appearance of conflicts which are weaker power-laws or with slopes away from 2.5 do not invalidate our model. 

\section{Conclusion}
We have provided three contributions which help unify disparate existing claims in the literature, and support the notion of a unified framework for understanding human conflict. We showed that the approximate power law obtained for whole war sizes bears a simple relationship to the approximate power laws obtained from events within individual wars. This connection that we find between event-level war data and war-level aggregated data, is new. We also clarified some misunderstandings relating to the application of power-law estimation and testing procedures to the messy conflict datasets available to modern researchers. 
Finally, we discussed how a generative theory of human conflict can be provided that reproduces the observed casualty distributions but does not make any implicit assumptions about power-law behavior. While it can indeed reproduce approximate power-law behaviors, it also provides a quantitative explanation of departures of these distributions from pure power laws in terms of the dynamics of the conflict.

\let\oldbibliography\thebibliography
\renewcommand{\thebibliography}[1]{\oldbibliography{#1}
\setlength{\itemsep}{1pt}} 
\bibliographystyle{chicago}
\bibliography{references}

\end{document}